\documentclass[a4paper]{jpconf}
\usepackage{graphicx}
\begin{document}
\title{Self-similarity of strangeness production \\
 in $pp$ collisions at RHIC}

\author{M Tokarev$^1$ and I Zborovsk\'{y}$^2$}

\address{$^1$ Joint Institute for Nuclear Research, Dubna, Moscow region,  Russia}

\address{$^2$ 
Nuclear Physics Institute,
Academy of Sciences of the Czech Republic, 
\v {R}e\v {z}, Czech Republic}

\ead{tokarev@jinr.ru}
\ead{zborov@ujf.cas.cz}

\begin{abstract}
New experimental data on transverse momentum spectra of strange 
particles  $(K_S^0, K^-,K^*,\phi,...)$ produced in $pp$ collisions 
at $\sqrt s =200$~GeV 
obtained by the STAR and PHENIX collaborations at RHIC are analysed 
in the framework of $z$-scaling approach.
Scaling properties of the data $z$-presentation are illustrated.
Self-similarity of strange particle production 
is discussed.
A microscopic scenario of constituent interactions 
developed within the $z$-scaling approach is used to study 
constituent energy loss, proton momentum fraction  and recoil mass 
in dependence on the transverse momentum, strangeness, and mass 
of the inclusive particle.
The obtained results can be useful for understanding 
strangeness origin,   
for searching for new physics with strange probes  and can serve 
as a benchmark for complex analyses of self-similar features 
of strange production in heavy ion collisions. 
\end{abstract}

\section{Introduction}

The production of particles with high transverse momenta
from the collisions of hadrons and nuclei at high energies 
has relevance to constituent interactions at small scales. 
One of the leading principles governing these interactions 
is self-similarity. 
This principle reflects hadron structure at different scales, 
interaction of hadron constituents 
and processes of fragmentation into colorless particles. 
Strange particles as well as other probes
(direct photons, jets, lepton pairs) 
play an important role to study fundamental features 
of the produced matter in hadron and nuclear interactions 
and provide information 
about its transition into observed particles. 

The scaling behavior of particle production related to self-similarity 
of hadron interactions at constituent level is manifested by 
the $z$-scaling \cite{1,2}.  
The concept of this scaling 
was used for analysis of inclusive spectra obtained 
at the accelerators  U70, $\rm S\bar p pS $, SPS, ISR, Tevatron and RHIC 
\cite{3}-\cite{9}.
The experimental spectra reveal striking similarity 
over a wide range of kinematic variables when expressed by the variable $z$. 
The $z$-scaling is treated as a manifestation of the self-similarity 
of the structure of the colliding objects,
interaction mechanism of their constituents, 
and processes of fragmentation into real hadrons. 
The approach can be a suitable tool to search 
for phase transitions and the critical point in hadron and nuclear matter. 
The parameters of the $z$-scaling, $c,\delta$ and $\epsilon$, 
have physical interpretation 
as the heat capacity of the produced matter, the fractal dimension 
of the structure of hadrons or nuclei and the fractal dimension 
of the fragmentation process, respectively.
Universality of the scaling is given by its flavour independence.
It means that spectra of particles with different flavour content 
can be described by a universal function $\Psi(z)$ when using
the scale transformation   
$z\rightarrow \alpha_F z$, $\Psi \rightarrow \alpha_F^{-1} \Psi $.

In this contribution we illustrate properties of the $z$-presentation 
of the transverse momentum distributions of strange particles
produced in $pp$ collisions. 
We present results of new analysis of data on inclusive cross sections
obtained by the STAR and PHENIX collaborations at RHIC. 
The dependence of the momentum fractions  and recoil mass 
on the collision energy, transverse momentum  and mass 
of the inclusive particle is used to estimate the constituent energy loss.

\section{$z$-Scaling}

The $z$-scaling has been suggested to express general regularities found
in the inclusive hadron production in high energy proton-(anti)proton 
and nucleus-nucleus collisions.
It manifests itself in the fact that the inclusive spectra of various types of particles can be described
with a universal scaling function. The function $\Psi(z)$  depends on a single variable 
$z$  in a wide range of the transverse momentum, registration angles, collision energies and centralities.
The scaling variable is expressed by the formula:
\begin{equation}
z=z_0 \cdot \Omega^{-1}.
\label{eq:1}
\end{equation}
Here $z_0$  and $\Omega$ are functions of kinematic variables:
 \begin{equation}
z_0 = {{\sqrt{s_{\bot}}\over {(dN_{ch}/d\eta|_0)}^cm_N}  }
\label{eq:2}
\end{equation}   
\begin{equation}
\Omega = (1-x_1)^{\delta_1}(1-x_2)^{\delta_2}(1-y_a)^{\epsilon_a}(1-y_b)^{\epsilon_b}.
\label{eq:3}
\end{equation}
The variable  $z$ is proportional to the transverse kinetic energy 
of the selected binary constituent sub-process required for  production
of the inclusive particle with mass $m$  
and its partner (antiparticle). The multiplicity density $dN_{ch}/d\eta|_0$  
of charged particles in the central region, the nucleon mass $m_N$  
and the parameter $c$  completely determine 
the dimensionless quantity $z_0$. 
The parameter $c$  has meaning of the "specific heat" 
of the medium produced in the collisions.

The quantity  $\Omega $  is proportional to the relative number 
of the constituent configurations which include 
the binary sub-processes corresponding to the momentum fractions $x_1$
and $x_2$  of colliding hadrons (nuclei) and to the momentum fractions
$y_a$  and $y_b$  of the secondary objects produced in these sub-processes. 
The parameters $\delta_1$  and $\delta_2$  are fractal dimensions 
of the colliding objects, and $\epsilon_a$  and $\epsilon_b$ 
stand for the fractal dimensions of the fragmentation process 
in the scattered and recoil direction, respectively. 
For unpolarized processes we assume the later to have the same value  
$\epsilon_a=\epsilon_b=\epsilon_F$
which depends on the type of the inclusive particle.
The selected binary sub-process, which results in production 
of the inclusive particle and its recoil partner (antiparticle), 
is defined by the maximum of $\Omega(x_1,x_2,y_a,y_b)$ 
with the kinematic constraint
\begin{equation}
(x_1P_1+x_2P_2-p/y_a)^2=M_X^2.
\label{eq:5}
\end{equation}
Here  $M_X=x_1M_1+x_2M_2+m/y_b$ is the mass of the recoil system in the sub-process.
The 4-momenta of the colliding objects and the inclusive particle are $P_1, P_2$  
and $p$, respectively.
The microscopic scenario of constituent interactions developed 
in the framework of  $z$-scaling is based on dependencies of the momentum fractions 
on the collision energy, transverse momentum and centrality.
The scaling variable $z$  has property of a fractal measure. 
It grows in a power-like manner with the increasing resolution 
$\Omega^{-1}$ defined with respect to the constituent sub-processes which satisfy 
(\ref{eq:5}). 
The scaling function $\Psi(z)$ is expressed in terms 
of the inclusive cross section 
$Ed^3\sigma / dp^3$, multiplicity density $ dN/d\eta$, 
and total inelastic cross section $\sigma_{in}$. 
All these quantities are measurable for the inclusive reaction 
$P_1+P_2\rightarrow p+X$. The function $\Psi(z)$ is determined 
by the expression:
 \begin{equation}
\Psi(z)=-{{\pi}\over {(dN/d\eta)\sigma_{in}}} J^{-1} E {{d^3\sigma}\over {dp^3} }.
\label{eq:6}
\end{equation}
Here $J$ is Jacobian of the transformation from the variables 
$\{p_T^2,y\}$  to $\{z,\eta \}$.
The function $\Psi(z)$ is normalized as follows:
\begin{equation}
\int_0^{\infty} {\Psi(z)dz} = 1.
\label{eq:7}
\end{equation}			
Equation (\ref{eq:7}) allows us to interpret  $\Psi(z)$ as probability 
density of the production of the inclusive particle 
with the corresponding value of the variable $z$.

\section{Scaling properties of strangeness production}

The self-similarity of production of different hadron species 
in $pp$ collisions provides the basis for analysing scaling 
properties of the hadron spectra for more complicated systems. 
The hadrons with strange quarks constitute a special interest 
as they are the lightest particles with quantum numbers 
which are absent in the net amount in the initial state. 

\begin{figure}[b]
\begin{center}
\includegraphics[width=14pc,height=14pc]{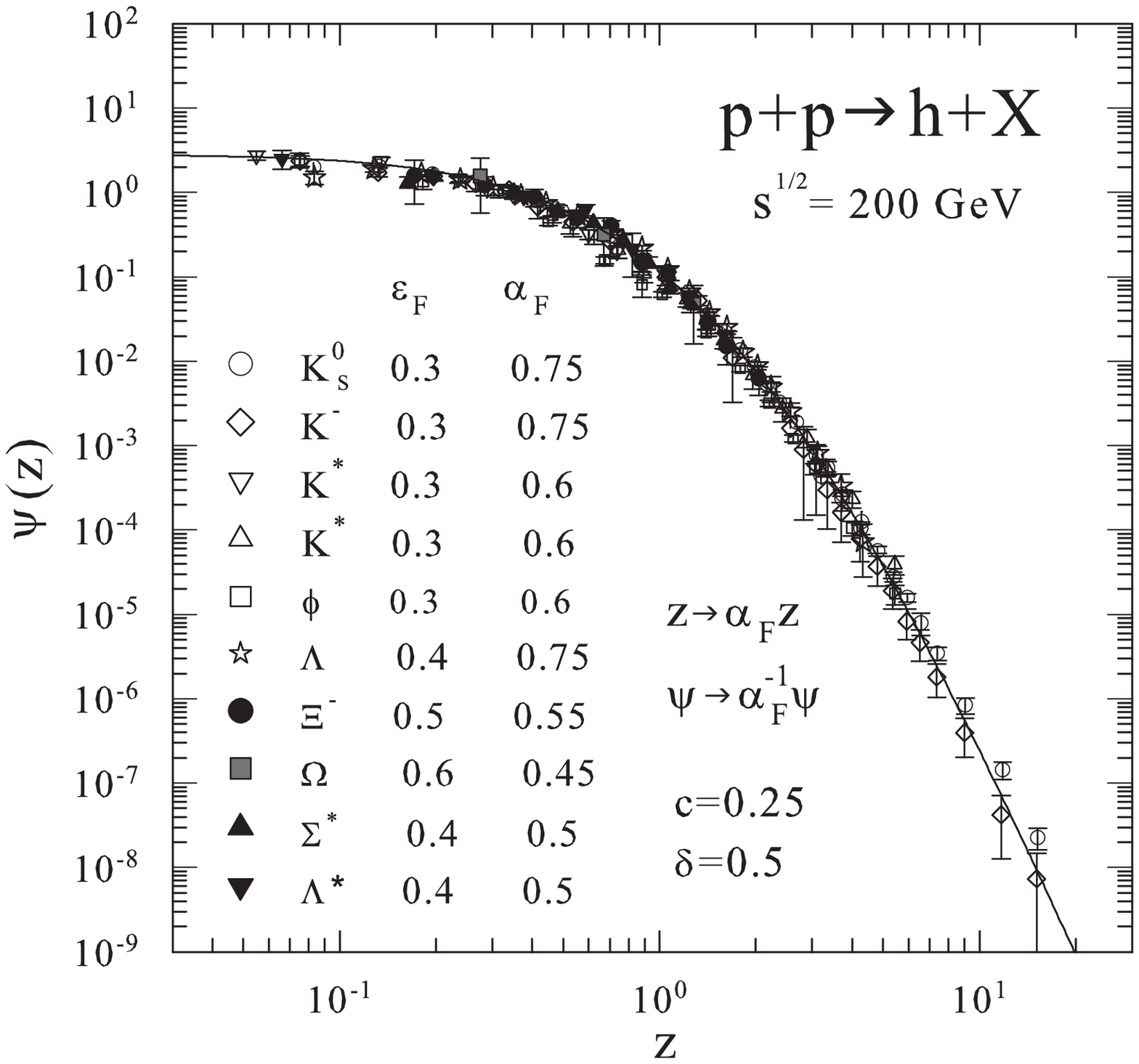}
\hspace*{10mm}
\includegraphics[width=14pc]{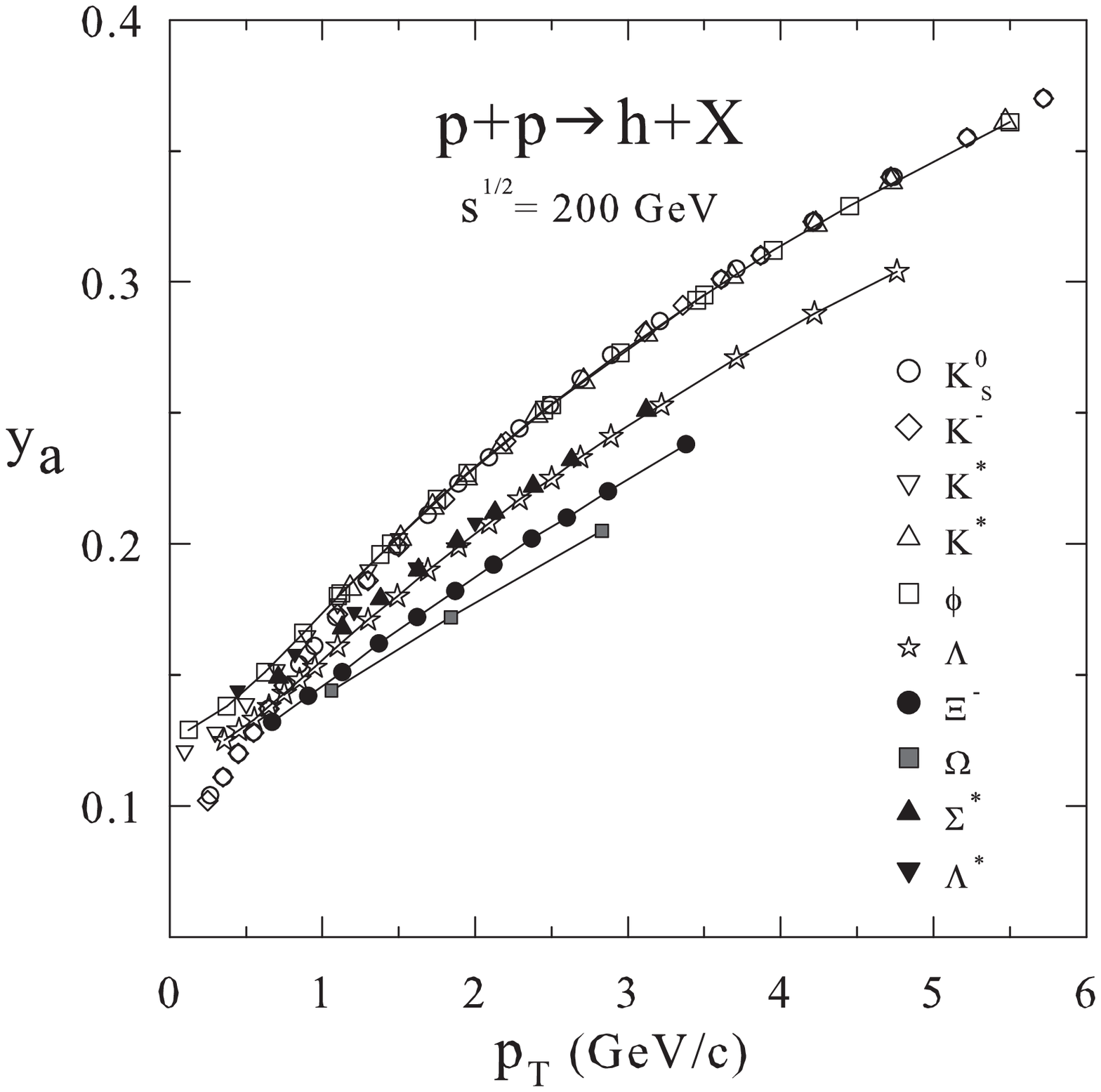}

\hspace*{10mm} a) \hspace{65mm}  b)
\end{center}
\caption{\label{label} (a) Inclusive transverse momentum spectra 
of strange particle production in $pp$ collisions in $z$-presentation.
(b) Relative energy loss 
$\protect{\Delta E/E = (1-y_a)}$ in dependence on $\protect{p_T}$.}
\end{figure}

Figure 1(a) shows $z$-presentation of the transverse momentum 
spectra of strange mesons and baryons measured 
in $pp$ collisions in the central rapidity region at RHIC. 
The symbols representing data on differential cross sections
 include baryons which consist of  one, two and three strange
  valence quarks. The open circles and diamonds correspond 
  to the respective spectra of $K^0_s$  and $K^-$ mesons 
   measured by the STAR collaboration 
\cite{10,11}.
%\cite{K0s_1  Km_1  Lambda  Xi Omega,K0s_2 and Km_2}.
 The data on the neutral short-lived $ K^*(892)$ resonance obtained by the STAR \cite{12}
% \cite{K*_ST} 
 and PHENIX \cite{13}
% \cite{K*_PX}
  collaborations are depicted by the triangles down and triangles up, respectively. 
The open squares correspond to the PHENIX data on 
$ \phi $-meson production \cite{14} 
%\cite{phi_PX} 
 detected in the $ K^+K^- $ and $ e^+e^- $  decay channels.
The $z$-presentation  of spectra of strange baryons shown in the figure is based on data collected by the STAR collaboration \cite{10}. 
%\cite{K0s_1  Km_1  Lambda  Xi Omega}. 
The distributions of $\Lambda$, $\Xi^-$ and $\Omega$  baryons 
are depicted by stars, black circles and black
squares, respectively.
The spectra of strange baryon resonances $\Sigma^*(1385)$ (full triangles up)
and $\Lambda^*(1520)$ (full triangles down) were taken from \cite{15}.
%\cite{Baryon res_ST}. 

The solid line represents the scaling curve obtained from analysis of pion production
in $ pp/\bar{p}p$ collisions ($\epsilon_{\pi}=0.2$, $\alpha_{\pi}=1$)
 over a wide range of kinematic variables. 
The curve is consistent with the energy, angular and multiplicity
 independence of the scaling function for different hadrons.
One can see that the corresponding fragmentation dimension
 $ \epsilon_F $ for strange mesons is larger than for pions,
  suggesting larger energy loss by production of mesons with
   strangeness content.  The fragmentation dimension 
   for strange baryons grows with the number of the 
    strange valence quarks. This is connected with the increase of energy losses.

The $p_T$-dependence of the energy loss by production
 of strange hadrons is shown in Fig.1(b). 
 The relative energy loss $ \Delta E/E = (1-y_a)$ depends on value
  of the fragmentation dimension $\epsilon_F$. 
As one can see, the relative energy loss decreases 
with increasing transverse momentum for all particles.
For particles with a given $ p_T>1$ GeV/c, 
the energy loss is larger for strange baryons than 
for strange mesons. The growth indicates increasing 
tendency with larger number of strange valence quarks inside the strange baryon.   
Such a tendency corresponds to the $z$-scaling universality (Fig.1(a))
 for hadron production including strange mesons and baryons.   

\begin{figure}[h]
\begin{center}
\includegraphics[width=14pc]{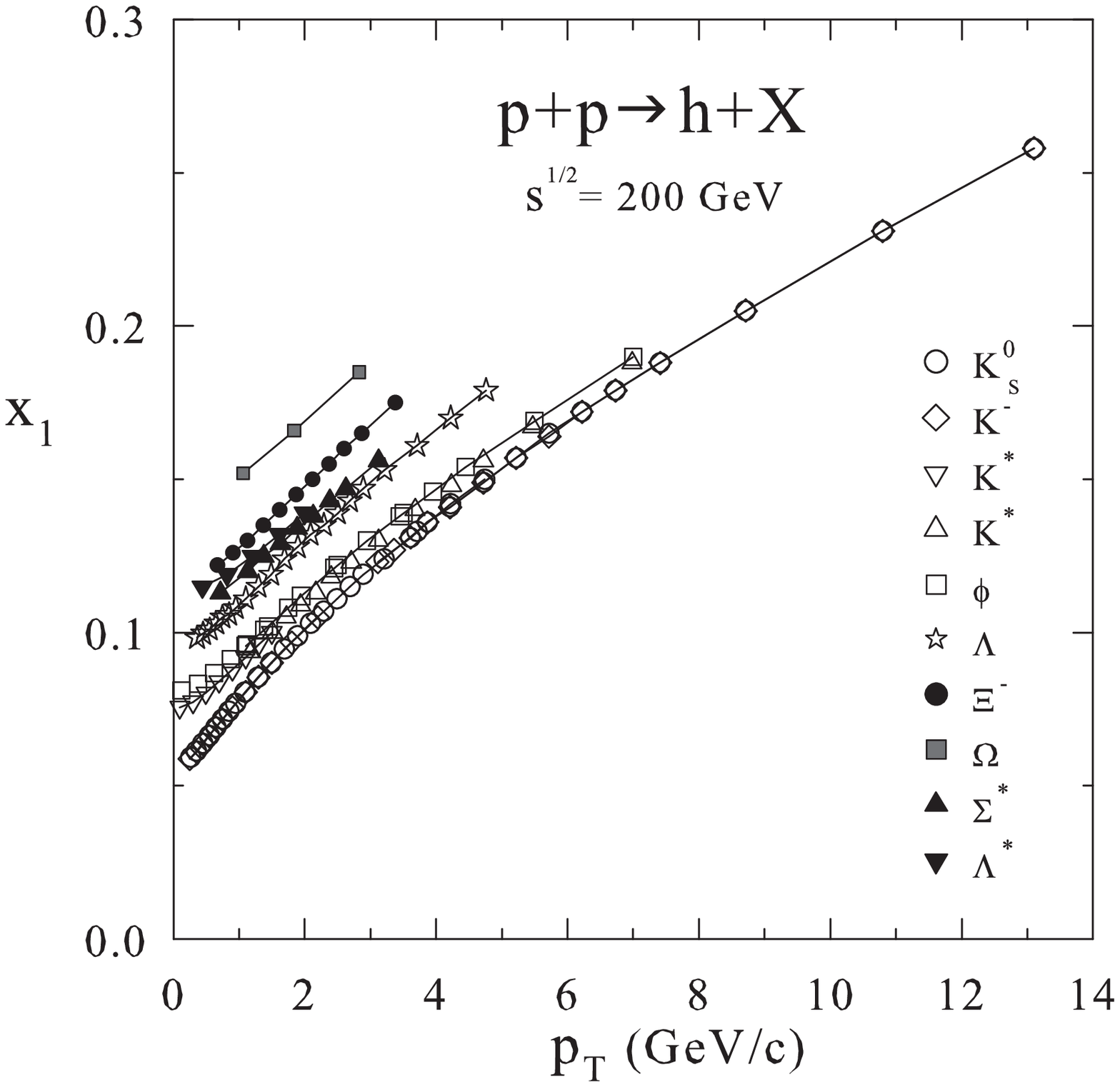}
\hspace*{10mm}
\includegraphics[width=14pc]{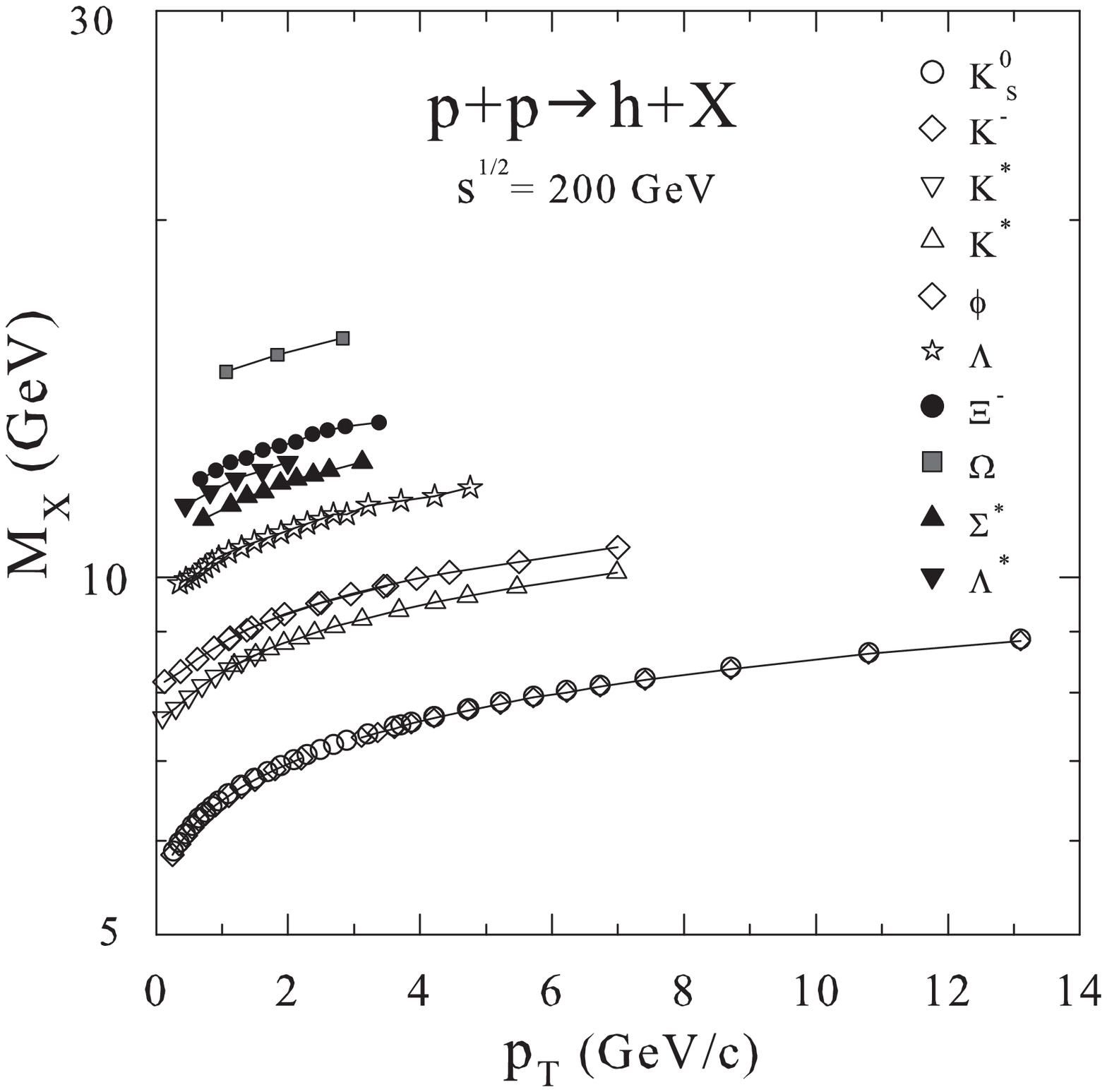}

\hspace*{10mm} a) \hspace{65mm}  b)
\end{center}
\caption{\label{label}  The $\protect{p_T}$-dependence of the momentum fraction 
$\protect{x_1}$  (a) of the colliding proton and the recoil mass $\protect{M_X}$ (b)
for production of strange hadrons.  }
\end{figure}

Figure 2(a) shows $p_T$-dependence of the momentum fraction
 $ x_1 $ for production of different hadrons. The fraction
  $ x_1 $ characterizes amount of energy (momentum) 
  of the incoming proton carried by the interacting constituent 
  which underlies production of the inclusive particle with momentum  $p_T$. 
One can see that, for given strangeness and $ p_T$, the value of 
 $x_1$ increases with the mass of the strange meson. 
The same tendency is observed for the recoil mass $M_X$ (Fig.2(b)).

In summary, we have analysed new transverse momentum spectra 
of strange particles measured by the STAR and PHENIX collaborations
 in $pp$ collisions at $ \sqrt{s} $=200 GeV using data $z$-representation. 
The analysis confirmed flavor independence of 
the scaling function $\Psi(z)$. The universality of
 description of the spectra is consistent with a hierarchy 
 of the energy loss in dependence on hadron strangeness and
  for a given strangeness with the increasing tendency 
  of the recoil mass $M_X$ and the constituent energy 
  with the mass of the strange hadron.

\smallskip


\medskip
\begin{thebibliography}{9}
%1
\bibitem{1} 
Zborovsk\'{y} I and Tokarev M V 2007 {\it Phys. Rev.} D {\bf 75} 094008

%2
\bibitem{2} 
Zborovsk\'{y} I and Tokarev M V 2009 {\it  Int. J. Mod. Phys.}  A {\bf 24} 1417

%3
\bibitem{3}
Tokarev M V 2006 {\it Phys. Part. Nucl. Lett.} {\bf 3} 7

%4
\bibitem{4}
Tokarev M V 2007 {\it Phys. Part. Nucl. Lett.} {\bf 4} 676

%5
\bibitem{5}
Tokarev M V and Zborovsk\'{y} I 2012
{\it Phys. At. Nucl.} {\bf 7} 700

%6
\bibitem{6}
Tokarev M V and Zborovsk\'{y} I 2010
{\it Phys. Part. Nucl. Lett.} {\bf 7} 171

%7
\bibitem{7}
Tokarev  M V (for the STAR Collab.) 2011  
{\it Phys. At. Nucl.} {\bf 74} 799

%8
\bibitem{8}
Tokarev M V and Zborovsk\'{y} I 2012
{\it J. Mod. Phys.} {\bf 3} 815

%9 
\bibitem{9}
Tokarev M V and Zborovsk\'{y} I 2013
{\it Nucl. Phys. Proc. Suppl.} B {\bf 245}  231
 
% 10 K0s_1  Km_1  Lambda  Xi Omega 
\bibitem{10}
Abelev B I {\it et al} 2007 {\it Phys. Rev.} C {\bf 75} 064901 

% 11 K0s_2 and Km_2
\bibitem{11}
Agakishiev G {\it et al} 2012 {\it Phys. Rev. Lett.} {\bf 108} 072302  

% 12  K*_ST 
\bibitem{12}
Adams J {\it et al} 2005 {\it Phys. Rev.} C {\bf 71} 064902

%  13 K*_PX 
\bibitem{13}
Adare A {\it et al} 2014 {\it Phys. Rev.} C {\bf 90} 054905 

% 14 phi_PX 
\bibitem{14} 
Adare A {\it et al} 2014 {\it Phys. Rev.} D {\bf 83} 052004 

%  15 Baryon res_ST 
\bibitem{15} 
Abelev B I {\it et al} 2006 {\it Phys. Rev. Lett.} {\bf 97} 132301  
\end{thebibliography}
\end{document}